\documentclass{aa}
\usepackage{graphicx}
\usepackage{color}
\begin{document}
   \title{Contribution of starburst mergers at $z\sim\,1$ to the strong
   evolution of infrared and submillimeter deep surveys}


   \author{Y.P. Wang}

   \offprints{Y. P. Wang}

   \institute{Purple Mountain Observatory, Academia Sinica, China; National Astronomical Observatories of China\\
              National Astronomical Observatory of Japan, Mitaka, Tokyo 181-8588, Japan\\
              \email{ypwang@pmo.ac.cn} }

   \date{Received 20 July 2001; accepted 29 November 2001}

 \abstract{
   Recent far-infrared and submillimetre waveband observations
   revealed large number of Ultraluminous
   Infrared Galaxies (ULIGs) with infrared luminosities $> 10^{12}\,L_{\odot}$.
   These sources are
   proposed to lie at redshifts above one, and in normally interacting systems with
   very dusty environments. We discussed in a previous paper that a population with a
   fast evolving infrared burst phase triggered by gas-rich mergers
   at $z\sim 1$ interpreted successfully the steep slope of faint IRAS
   $60\,\mu m$ source counts within the flux range of $100\,mJy \sim 1\,Jy$, still leaving
   the infrared background level at this wavelength compatible with the upper limit from recent
   high energy TeV $\gamma$ ray detection of Mrk501. To extend the model to mid and far infrared wavelengths, we adopt a reasonable
   template spectral energy distribution as typical for nearby infrared bright starburst galaxies ($L_{ir} <= 10^{12}\,L_{\odot}$),
   such as Arp220. We construct the SED
   for the dusty starburst mergers at $z\sim 1$ by a simple dust extinction law and a thermal continuum assumption for the far-infrared
   emission. Since the radiation process at mid-infrared for these starburst merging systems is still uncertain, we assume
   it is similar to the MIR continuum of Arp220, but modify it by the observed flux correlation of ULIGs from IRAS and ISOCAM
   deep surveys.
   We show in this paper that the strong evolution of the European Large Area ISO Survey (ELAIS)
   at $90\,\mu m$, ISO $170\,\mu m$ and the Submillimeter deep survey at $850\,\mu m$ could be sufficiently accounted for by
   such an evolutionary scenario, especially the hump of the ISOCAM $15\,\mu m$ source count around $0.4\,mJy$.
   From current best fit results, we find that the dust temperature of those extremely bright starburst merging system at $z\sim 1$
   would be higher than that of Arp220 for a
   reconciliation of the multi-wavelength infrared deep surveys. We thus propose
   that the infrared burst phase of dusty starburst galaxies or AGNs from gas-rich mergers
   at $z\sim 1$ could contribute significantly to the strong evolution of the IRAS
   $60\,\mu m$, the ISO $15\,\mu m$, $90\,\mu m$, $170\,\mu m$, as well as the SCUBA $850\,\mu m$ number
   counts, while being compatible with the current observational limits of cosmic infrared background and the redshift distributions.
   The major difference of our current model prediction is that we see a fast convergence of the differential number counts
   at $60\,\mu m$
   below $50\,mJy$, which is about a factor of two brighter than other model predictions. Future infrared satellites like Astro-F or
   SIRTF would give strong constraints to the models.
  }

   \titlerunning{Starburst mergers and the strong evolution of IR deep surveys}
   \maketitle{}{}

\section{Introduction}
There has been much progress in the study of extragalactic
evolution since far-infrared and submillimeter deep surveys
detected a significant population of Ultraluminous Infrared
Galaxies (ULIGs: $L_{ir} > 10^{12}L_{\odot}$) at high redshift ($z
\sim 1-4$) (Hughes et al. 1998, 2000, Blain et
al. 1999, Eales et al. 1999, Holland et al. 1999, Puget et al.
1999, Sanders 1999, Dole et al. 2000). The major interests of present research are
the nature and evolution of these sources. Due to the lack of high
resolution morphological studies, the origin of these faint SCUBA
sources are still not clearly understood. Local ULIGs most likely
arise from major galaxy mergers with high dust extinction for the
central starburst or AGN activities (Sanders \& Mirabel 1996,
Murphy et al. 1996, Surace et al. 1998, 2000, Scoville et al.
2000). Nevertheless, more than 50$\%$ of the high redshift
ultraluminous infrared objects and the optical counterparts of
ISOCAM HDF-N galaxies are suggested to show indication of galactic
interactions and merger signatures (Mann et al. 1997, Smail et al.
1999, Ivison et al. 1998, 2000, Sanders 1999). Meanwhile,
estimation of their spectral energy distributions suggests that
these galaxies are probably the high redshift counterpart of the
local ULIGs discovered by the IRAS deep survey (Barger et al.
1998, 1999, Smail et al. 1998, Trentham et al. 1999, Frayer et al.
2000). Although the exact mechanism for formation of these
interesting sources are not clear, there are certain reasons to
say that majority of them may be major mergers of gas-rich disks
accompanied by dust-shrouded nuclear starbursts or powerful Active
Galactic Nuclei (AGN). SMM J14011+0252, and ERO J164502-4626.4
(HR10) are two such candidates with their central activities
heavily hidden by dust extinction, and are suggested to be
consistent with the evolutionary track of mergers--starbursts/AGN,
probably elliptical galaxies in formation (Graham \& Dey 1996,
Cimatti et al. 1998, 1999, Frayer et al. 1998, 1999, Dey et al.
1999, Papadopoulous et al. 2001).

On the other hand, the source counts from present infrared and
submilimetre surveys, such as IRAS, ISO and SCUBA all
significantly exceed the non-evolving predictions. The extremely
strong evolution is seen from the differential counts of the
ISOCAM at $15\,\mu m$, with the remarkable upturn at $S_{15}<
3\,mJy$ and a fast convergence when $S_{15}\sim 0.3\,mJy$. This
striking feature is based on the data from several independent sky
surveys (Elbaz et al. 1999, Chary \& Elbaz 2001, Mazzei et al.
2001, Serjeant et al. 2001). Although there are many other
possible evolutionary scenarios which could explain the present
observations, the reason that we are encouraged to explore here a
merger-driven galaxy evolution picture with binary aggregation
dynamics is simply because the IRAS database and recent ISO and
sub-mm deep surveys indicate that most of the luminous infrared
sources are actually interacting/mergering systems. Also, the
local IR luminosity function shows an excess over the
Press-Schechter formula (Press \& Schechter 1974, Lonsdale 1995,
Pearson \& Rowan-Robinson 1996, Guiderdoni et
al. 1998, Roche et al. 1998, Rowan-Robinson et al. 1998, Dey et
al. 1999, Sanders 1999, Dole et al. 2000, Efstathiou et al. 2000, Silk \& Devriendt
2000, Serjeant et al. 2001, Takeuchi et al. 2001).

Considering mergers as a possible formation mechanism of
Ultraluminous Infrared Galaxies both at high and low redshift, as well as their
significant infrared emissions, Wang (1999) and Wang \&
Biermann (2000) discussed the effects of galaxy mergers on the
strong evolution of the IRAS $60\,\,\mu m$ deep survey within a
binary aggregation galactic evolutionary scheme. In this model, the bright tail of the infrared luminosity function is
simulated in a consistent way for both the density and luminosity
evolution due to the decrease of the merger fration with cosmic time and a merger-triggered infrared burst phase.
They found that a
luminosity-dependent infrared burst phase is crucial for the
interpretation of the steep slope within a flux range of
$10\,mJy\sim 1\,Jy$ by the IRAS $60\,\,\mu m$ deep survey. This means
dusty starburst galaxies or AGNs from gas rich mergers at high
redshift may experience an infrared burst phase around a
transition redshift $z\sim 1$, and fade quickly within the merger
time scale of that epoch. The more massive merger systems could
have such infrared emission enhanced to a higher level and
decrease even faster. This kind of speculation is based on the
observation that ULIGs are usually more than a
factor of 20 brighter than normal starburst galaxies. Although the
detailed mechanism for such enormous infrared emission is still
unclear, it is believed to be related to a special stage of the
merger process when the dust mass and temperature are both
dramatically increased (Kleinmann \& Keel, 1987, Taniguchi \&
Ohyama 1998). Recent numerical simulation on the evolution of
dusty starburst galaxies by Bekki \& Shioya (2001) shows that
there is a very strong photometric evolution during the merger
process of two gas-rich disks, and a dramatic change of the
spectral energy distribution (SED) around a cosmic time scale $T
\sim 1.3\, Gyr$, when the two disks of the merger become very
close and suffer from violent relaxation and the star formation
becomes maximal ($\sim 378\,M_{\odot}yr^{-1}$). The infrared flux
in this case could increase by one magnitude, especially for the
far infrared wavelength range ($60\,\mu m \,\sim 90\,\mu m$) in the
emitting frame.

The redshift distribution of the contributing sources for the
steep slope at faint IRAS $60\,\mu m$ counts in the model of Wang
\& Biermann (2000) shows that the infrared burst phase around
$z\sim 1$ could have comparable significance to the local IR
sources. The question is then whether such an
infrared burst phase, or such a population of ULIGs, could also
sufficiently account for the strong evolution seen in other infrared
wavelengths, especially at the ISOCAM $15\, \mu m$, ISOPHOT $90\,\mu m$, $170\,\mu m$ and
SCUBA $850\,\mu m$. We thus try to make a reconciliatory evolution model which could fit at least the present statistics of the
multi-wavelength deep surveys.

In this paper, we will first review the binary aggregation galaxy
evolution model by Wang (1999) and by Wang \& Biermann (2000) in section
2, where starburst/AGN activities may be triggered during the
merger process, as well as an infrared burst phase from gas-rich
mergers around a redshift of one. We will discuss the SED template we adopt in our calculation for
the nearby starburst galaxies and a possible strong evolution of the spectral energy distribution
of the dusty starburst merging systems
at $z\sim 1$. We thus could further investigate whether the infrared
burst phase from gas-rich mergers around redshift $z\sim 1$ is
sufficient to account for the strong evolution also detected by
ISO and submillimetre deep surveys. One set of cosmological parameters, namely
$H_{0}=50\,km/s/Mpc, \,\Omega=0.3$ and $\Lambda=0.7$ is adopted in the
calculation.

\section{Model}
We adopt in this study the binary aggregation dynamics based on
the Smoluchowski equation (1916) within a merger-driven galaxy
evolutionary scheme, which simulates the evolution of a luminosity function due to galaxy mergers
and predicts the redshift-dependent luminosity
function of a population of galaxies evolving
forwards in
time from the formation epoch to match the observed local
luminosity function, number counts and space distribution etc.
Studies of Cavaliere \& Menci (1993, 1997) show that this method
could include more dynamics to describe a further step in
galaxy-galaxy interactions within the scheme of direct
hierarchical clustering (DHCs), and probably could help to
alleviate some intrinsic problems in the DHC scenario, such as the
overproduction of small objects as well as the difficulty of a
reconciliation between the excess of faint blue counts and the flat
local luminosity function. The numerical technique to solve the
Smoluchowsky equation is a Monte-Carlo approach for the
inverse-cascade merger tree. Readers are referred to Cavaliere
\& Menci (1993, 1997), Wang (1999) and Wang \& Biermann (2000) for
the details of the dynamics and the techniques.

Considering different evolutionary characteristics of different
morphologies, we adopt in our study a multi-component model that
contains starburst galaxies, dust shrouded AGNs and spiral
galaxies as three major classes of infrared emitting sources. The
local luminosity functions of the spiral and starburst galaxies at
$60\,\mu m$ from Saunders (1990), and that of Seyferts from Ruch et al. (1993)
are used to normalize the Monte-Carlo simulation. We adopt the
mass-light relation of blue starburst galaxies, which is given by
Cavaliere \& Menci (1997) in a study of the excess of faint blue
galaxies in optical surveys. The abundances of dust-shrouded AGNs
are set to be $50\%$ and $80\%$ at local and high redshift based
on the statistics from Hubble Space Telescope imaging survey of
nearby AGNs and the cosmic X-ray background (Malkan et al. 1998,
Gilli et al. 1999).

The modelling of a luminosity-dependent ultraluminous infrared
burst phase from gas-rich mergers is described in detail by Wang
\& Biermann (2000). Here we give a brief review of the basic
dynamics and the template SED we adopt in our model for the infrared luminous sources.
We introduce also in this section the construction of a SED for the dusty starburst mergers at $z\sim 1$,
normally with the luminosity $L_{ir}> 10^{12}\,L_{\odot}$, in order to further
investigate such an evolutionary scenario at mid and far-infrared
wavelengths from the ISO deep survey. Considering that star formation is
triggered by mergers and proportional to $M_{gas}/\tau$ ($\tau$
is the dynamical interaction time scale), Cavaliere \& Menci
derived a mass-light ratio for dwarf galaxies, $L/L_{\ast} =
(M/M_{\ast})^\eta$ (where $\eta = 4/3$, if the cross section is
purely geometrical). $L_{\ast}\propto f(z,\lambda_{0},\Omega_{0})$ could be used to
describe a redshift dimming, or a luminosity evolution. A power law prescription of $f(z)\propto (1+z)^
{\beta}$
is adopted in the model. Simplifying the color and K-
corrections, they roughly get $L_{B}\propto
\frac{L_{\ast}}{M_{\ast}^{\eta}}\,M^{\eta}=
\frac{L_{\ast}(0)}{M_{\ast}^{\eta}\,f(z)\,M^{\eta}}$. We assume in
our model a luminosity ratio $\frac{L_{60}}{L_{B}}\propto
M^{\acute{\eta}}$, which is consistent with current understanding
of the nature of ULIGs, where people normally believe that the
extremely infrared bright phase is due to the starburst merger
events with the far-infrared luminosity $L_{ir}$ enhanced both by the
accumulation of the dust mass and the increase of the dust
temperature. This burst phase could enhance the infrared
luminosity by a factor of about 20 over that of normal starburst
galaxies (Kleinmann \& Keel, 1987, Taniguchi \& Ohyama 1998, Bekki
\& Shioya 2001). We give $L_{60}\propto
\frac{L_{\ast}(0)}{M_{\ast}^{\eta}}\,f(z)\,M^{\eta+\acute{\eta}}$.
$\acute{\eta}=1\sim\,1.2$ is adopted in the calculation, which not
only reasonably represents an infrared enhancement of about a
factor of 20 for a typical ULIG with a mass of $10^{12}\,M_{\odot}$
(the mass increases about one magnitude over that of normal starburst
galaxies), but successfully interprets the steep slope of IRAS
number counts. The scaling factor of the mass-light ratio here is
normalized by the local luminosity function of the IRAS deep survey. We found from our best fit results that a population of
infrared starburst sources, especially with spheroidal morphology, would have experienced very strong evolution
in the past. A rate of $\beta=3.7$ in the luminosity evolution
$f(z)\propto\,(1+z)^\beta$ since a transition redshift $z\sim 1$ indicates a very strong evolution for such a population of starburst
galaxies. This is at least comparable to, if not stronger than, QSOs (Roche et al. 1998, Lonsdale 1995, Pearson \&
Rowan-Robinson 1996, Rowan-Robinson et al. 1998,
Sanders 1999, Dole et al. 2000, Franceschini et al. 1988, 2001). A differential dimming is simulated by
$L_{ir}(z-\delta z) \propto L_{ir}(z)^{1-\zeta}$ below a
transition redshift $z\sim\,1$, in order to match the observed
local luminosity function by the IRAS deep survey. The simulation
gives a best value $\zeta \sim\,0.4$. This power law
suppression also includes another physical reality, that the infrared
luminous galaxies at the bright tail of the luminosity function
become gas poor faster than the less luminous ones. Besides the merger rate decrease with cosmic time,
this physical effect is very important for a good fit of the steep
slope at IRAS $60\,\mu m$ number counts within a flux range of
$100\,mJy\,\sim 1\,Jy$.

We reviewed above the dynamics and some important physical
parameters in the current study, which are the same as those used
in the previous paper of fitting IRAS $60\,\mu m$ number counts
(Wang 1999, Wang \& Biermann 2000). In the following, we will
start to construct the spectral energy distribution for dusty
starburst mergers around redshift $z\sim\,1$, in order to
extrapolate the calculation to mid- and far-infrared wavelengths.
Because the unclear nature of the Ultraluminous Infrared Galaxies
at high redshift, we do not have a good understanding of the dust
environment and properties in these sources. An optically thin,
single-temperature dust model is adopted as a first order
approximation for a modified blackbody continuum of temperature
$T$ at far-infrared wavelength in this calculation. The formula is
simply given by $S_{\lambda}=B_{\lambda}(T)\,\tau_{\lambda}\propto
B_{\lambda}\,K_{\lambda}$. $\tau_{\lambda}=K_{\lambda}\,\rho\,dl$
is the dust opacity, and $K_{\lambda}$ the dust absorption
coefficient ($K_{\lambda}\propto \lambda^{-\beta}$ of $\beta
\simeq 1-2$). In this case, the flux received at wavelength
$\lambda$ is
$S_{\lambda}=\frac{\Gamma\,\lambda_{e}^{-\beta}\,B(\lambda_{e},
T)}{4\,\pi\,dL^2\,(1+z)}$, where $\Gamma$ is the scaling factor
for a conservation of the dust absorbed energy and re-emitting
energy, $\lambda_{e} =\lambda/(1+z)$ is the wavelength in the
emitting frame. A flux ratio in the observer's frame could be
derived by
$\frac{S_{\lambda_{1}}}{S_{\lambda_{2}}}=\frac{\lambda_{1e}^{-\beta}\,
B(\lambda_{1e},T)}{\lambda_{2e}^{-\beta}\,B(\lambda_{2e},T)} \sim
(\frac{\lambda_{2}}{\lambda_{1}})^{\beta +5}\,e^{\frac{h\,C}{k}\,
(\frac{1}{\lambda_{2}}-\frac{1}{\lambda_{1}})\,\frac{1+z}{T}}$,
where $h$ is Plank's constant, $k$ is Boltzmann's constant and $C$
is the speed of light. With a reasonable assumption for the dust
emissivity power $\beta$ and the dust temperature $T$, we can
easily extend our calculation to far-infrared wavelengths. The
mid-infrared emission is more complicated than that of the
far-infrared which could be well described by a single temperature
blackbody spectrum by cold, large grain dust. The MIR emission
properties are usually dominated by the radiation field of heated
small grains and PAHs. These dust grains are normally heated
stochastically, and are not in thermal equilibrium with the
ambient radiation field. Thus the MIR continuum is most like a
power law spectrum. In this calculation, we will not go to a
detailed modelling of the MIR emission feature. Instead, we only
modify the template SED of the starburst galaxy Arp220 by Silva et
al. (1998), with the observational correlations of $S_{15}/S_{60}$
by IRAS and ISOCAM deep surveys for the ultraluminous case
($L_{ir} > 10^{11}\, L_{\odot}$) to represent the dusty starburst
merging system around redshift one. The color ratio of
$S_{15}/S_{60}$ for the ULIGs are a factor of about 5 lower than
the mean value of the whole sample, which may imply a very
complicated process to heat small grains during the merger process
(Aussel et al. 2000, Dunne et al. 2000, Saunders et al. 2000,
Chary \& Elbaz 2001). Although there are many indications that the
MIR continuum is correlated with the temperature of large grain
dust (Dale et al. 2001), there is still no exact modelling for
such a process. Our goal in this paper is to construct a simple
SED for those luminous starburst mergers based on various
obsevational correlations, which not only represents the observed
trend for individual samples of a certain luminosity bin
($L_{ir}>10^{12}\,L_{\odot}$), but matches the statistical results
from the multiwavelength deep surveys. The number of sources $dN$
in a comoving volume $dV$ within the flux range $S_{\lambda}$ to
$S_{\lambda}+ dS_{\lambda}$, measured at wavelength $\lambda$, is
defined by:
$dN=\rho(L_{\lambda},z)\,dV\,\frac{dL_{\lambda}}{dS_{\lambda}}\,dS_{\lambda}$,
$\frac{dL_{\lambda}}{dS_{\lambda}}=\frac{4\,\pi\,d_{L}^2}{K(L_{\lambda},z)}$,
where
$K(L_{\lambda},z)=\frac{L_{\lambda_{e}}\,d\lambda_{e}}{L_{\lambda}\,d\lambda}$
is the K-correction, and $d_{L}$ is the luminosity distance in a
$\Lambda$ dominated universe.

\section{Results and discussions \label{discussion}}
The exact broad-band spectra of faint IR sources is still not well
defined. Considering deep surveys at various IR/submm wavelengths
would help to simultaneously constrain the evolution properties
and the typical spectral energy distribution of such sources. We
show in this section the comparison of our model prediction with
the ISOCAM $15\,\mu m$ survey data, IRAS $60\,\mu m$, ISOPHOT
$90\,\mu m$ and $170\,\mu m$ (FIRBACK Survey), as well as the SCUBA $850\,\mu m$
data. We also calculated the redshift distribution of these
sources within a certain flux range and the cosmic infrared
background level. Fig.~\ref{fig1} to Fig.~\ref{fig3} are the model
predictions for the European Large Area ISO Survey (ELAIS). This
survey covered 12 $deg^2$ of the sky in four main areas and was
carried out with the ISOPHOT instrument onboard the Infrared Space
Observatory ISO, which is at least an order of magnitude deeper
than the IRAS $100\,\mu m$ survey. It therefore provides an
important constraint for our model of galaxy evolution. The
majority of the optical identification of the detected sources are
for interacting pairs or small groups of galaxies, which may
indicate that the ELAIS sample includes a significant fraction of
luminous infrared galaxies from galaxy mergers. Although there is
some discrepancy in the data reduction, previous estimations show
that the source counts are mostly in agreement with strongly
evolving starburst models, with a rapid increase in the fraction
of ULIGs towards high redshift (Efstathiou et al. 2000, Matsuhara
et al. 2000, Serjeant et al. 2001). From our calculations, we see
in Fig.~\ref{fig1} to Fig.~\ref{fig3} that the differential number
counts of $90\,\mu m$, $15\,\mu m$ and $170\,\mu m$ for a reliable
subset of the detected sources could be sufficiently accounted for
by the infrared burst phase when a population of ultraluminous
infrared sources with $L_{ir}>10^{12}\,L_{\odot}$ could be
produced by the merger-triggered starburst/AGN activities at
$z\sim 1$ (Kawara et al. 1998, Elbaz et al. 1999, Efstathiou et
al. 2000, Dole et al. 2001). The enormous infrared emission,
especially at far-infrared wavelengths is modelled by a modified
black body spectrum which we discussed in the previous section. We
assume that the starburst merging system has a similar MIR
emission feature as Arp220, but modified by the observed flux
correlation of $S_{15}/S_{60}$ from IRAS and ISOCAM deep surveys.
We found from the calculation that the dust temperature of these
starburst merging systems would be higher than that of the nearby
starburst ULIGs Arp220, with dust temperature $T= 65\,K$ and
$\beta = 1.5$ for a best fit result. Fig.~\ref{fig4} shows our
model fitting for the differential counts of the IRAS $60\,\mu m$
deep survey, and Fig.~\ref{fig5} shows the integrated number
counts of submillimeter SCUBA deep survey at $850\,\mu m$ (Hacking
et al. 1987, Moshir et al. 1992, Barger et al. 1999, Blain et al.
1999). Almost all the number counts could be reproduced quite well
by such an evolutionary scenario, except for the ISOCAM $15\,\mu
m$ differential number counts, where our model prediction shows a
slight excess at the bright part of $S_{15}\sim 2\,mJy$. The
reason could be that we simply adopt the mid-infrared emission
feature of Arp 220 for the case of the starburst merging system
around $z\sim 1$. We hope to improve the current results by
further theoretical modelling and additional observational
constraints for the emission properties at mid infrared bands from
future infrared missions.

The infrared background in this calculation gives 2.4$nW\,m^{-2}\,sr^{-1}$ at 15$\mu m$,
1.9$nW\,m^{-2}\,sr^{-1}$ at 60$\mu m$, 3.8$nW\,m^{-2}\,sr^{-1}$ at 90$\mu m$, 10.6$nW\,m^{-2}\,sr^{-1}$
at 170$\mu m$, which are all consistent
with current upper limits from TeV detections, COBE results and the resolved fraction of the CIRB by the deep ISO
surveys (Funk et al., 1998, Guy et al. 2000, Hauser \& Dwek 2001).

The redshift distribution of the ISOCAM 15$\mu m$ contributing
sources within the detected flux range ($0.1\,mJy\sim 10\,mJy$)
from our model calculation is shown in Fig.~\ref{fig6}. It gives a
rough statistical finding that these luminous infrared sources
cover a wide redshift range of $0.5\sim 2.5$, peaking at $z\sim
1$. Comparing our model prediction and the redshift distribution
of 15$\mu m$ sources with $S_{15}>120 \mu Jy$ in the HDF North and
the z-distribution of sources in the CFRS field (Flores et al.
1999, Cohen et al. 2000, Aussel et al. 2001), we found that the
starburst mergers at $z\sim 1$ in our model are good candidates
for a strongly evolving population that results in the strong
evolution in mid- and far-infrared deep surveys. Recent redshift
estimation from sub-mm follow up of 10 known FIRBACK 170$\mu m$
ISO sources by Scott et al. (2000) suggests that they are in a
redshift range of $0\sim 1.5$, still consistent with our current
model predictions. However, these redshift determinations strongly
depend on the assumption of the dust properties. We need further
accurate measurements for the robust constraints of the models. We
discussed in a previous paper that shifting the peak redshift of
these ULIGs by a factor of 2 could affect the source count fitting
of the IRAS 60$\mu m$ deep survey, especially for a low redshift
peak ($z<0.5$). Strong evolution of the ULIGs to $z\sim 1$ may be
the most reasonable case for the existing model constraints from
both the infrared deep surveys and the cosmic infrared background
upper limits from high energy TeV detections, as well as the
indicated star formation history by UV/optical deep surveys (Lilly
et al. 1996, Connolly et al. 1997, Madau et al. 1998).

We plot the redshift distribution of ULIGs ($\nu
L_{\nu}>10^{12}\,L_{\odot}$) to understand the evolutionary
properties of the ULIGs from mergers in our model. A rapid
increase in the number density of ULIGs up to $z\sim 1$ is seen in
Fig.~\ref{fig7}, which is actually consistent with a scenario
where galaxy merger rates increase dramatically during that epoch
as seen from various observations and theoretical considerations
(Zepf \& Koo 1989, Carlberg 1992, Burkey et al. 1994, Carlberg et
al. 1994). However, the number density of ULIGs decreases beyond
$z\sim 2-3$, which may reflect a stage when merger pairs are
mostly dwarfs and the infrared emissions are less than
$10^{12}\,L_{\odot}$ even with intensive starburst activities
triggered by mergers. In this scenario, an infrared luminous tail
of the luminosity function may form at $z\sim 1$, with enormous
infrared emission enhancement.

There is still no firm statistical basis for the classification of
starbursts and AGNs from current spectroscopies. We thus adopt the observed
AGN local luminosity function of Rush et al. (1993) as a model constraint, and assume in
our calculations that the observed starburst galaxies and
Seyferts follow the same evolutionary track, based on naive
thinking that starburst/AGN may both be triggered by galaxy
interactions. We know the subtle
differences in the dust emission properties could result in a
different fraction of their contribution. This is still far too
early to discuss here. We thus adopt only the SED of the
Cloverleaf quasar which represents a phase poor in cold gas, as well as the dust enshrouded phase of F10214+4724
as two typical AGN templates in
our calculation. We know from the result that the AGN
contribution is only a small fraction of the whole and our current model
prediction is within the present understanding of
this issue, i.e. the starburst powered ULIGs are dominating over
the AGN powered ones (Fig.~\ref{fig7}) and may take over at higher
redshifts and in the higher luminosity case (Lutz et al. 1998, Tran et al.
2001).

   \begin{figure}
   \label{fig1}
   \includegraphics[height=8cm,width=8cm]{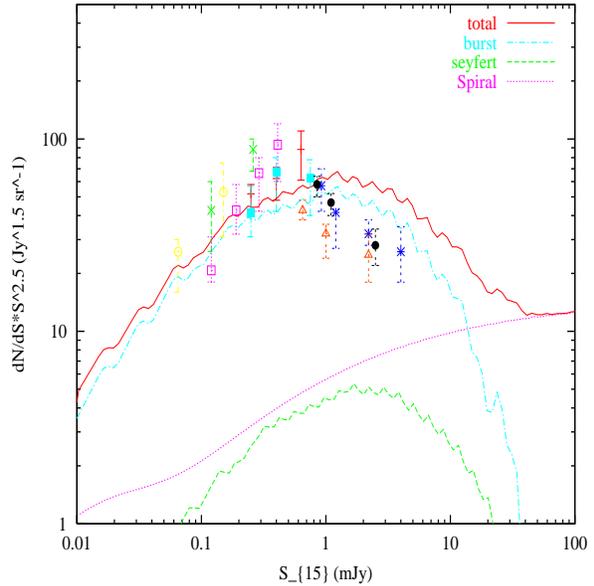}
   \caption{The model prediction of the differential number counts of ISOCAM $15\,\mu m$ normalized to
   the Euclidean law ($dN/dS\,*\,S^{2.5}$). The data points are the normalized counts from a variety of
   ISO deep surveys (Elbaz et al. 1999). The line represents the sum of the contribution from three populations
   (starburst galaxies, spiral galaxies and Seyferts). In our model, we assume that the population of starburst galaxies,
   especially with spheroidal morphology, are the product of galaxy mergers which would experience infrared
   emission enhancement because of the merger-triggered starburst activities. Their contribution to the ISOCAM $15\,\mu m$
   deep survey is shown by the dot-dashed line from our Monte-Carlo simulation.  The dotted line corresponds to
   the non-evolving spiral galaxies, and the short dashed line is from Seyferts, which are assumed to have the same
   evolutionary track as starburst galaxies in a galaxy evolutionary scheme with galaxy mergers.\label{fig1}}
   \end{figure}

   \begin{figure}
   \label{fig2}
   \includegraphics[height=8cm,width=8cm]{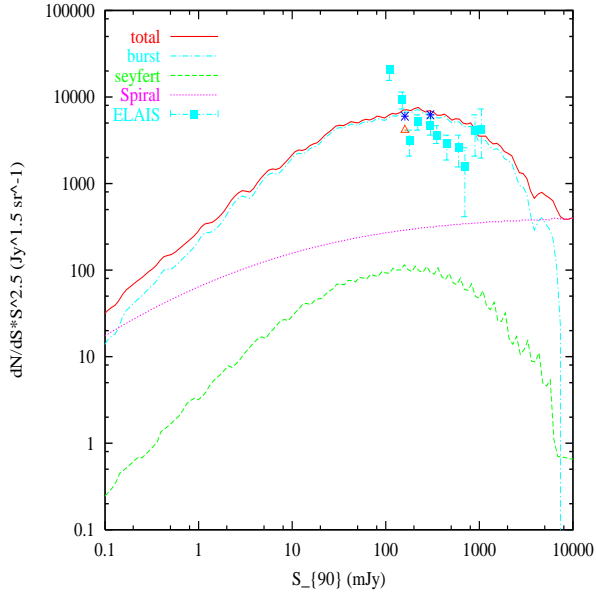}
   \caption{The fitting of the ELAIS
   differential source count at $90\,\mu m$. The data are from C-90 filter of the C100 ISOPHOT detector array [filled squares:
   Efstathiou et al. 2000; open trianguler: Linden-V$\phi$rnle et al. (2000); star: Juvela et al. (2000)].
   The meaning of the lines is the same as in Fig.\ref{fig1} \label{fig2}}
   \end{figure}

   \begin{figure}
   \label{fig3}
   \includegraphics[height=8cm,width=8cm]{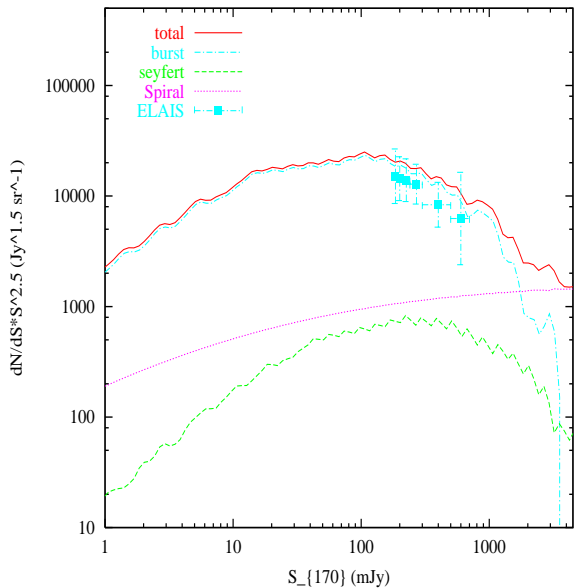}
   \caption{The result of FIRBACK $170\,\mu m$ ISO deep survey differential number count fitting from our model calculation.
    The data are from Dole et al. (2001). \label{fig3}}
   \end{figure}

   \begin{figure}
   \label{fig4}
   \includegraphics[height=8cm,width=8cm]{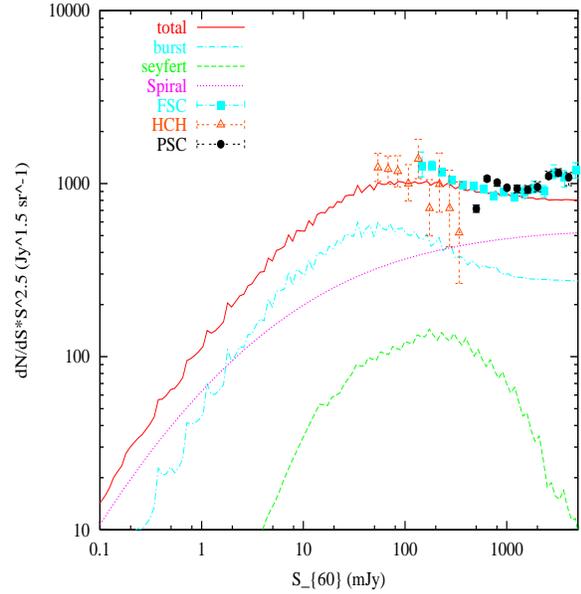}
   \caption{The fitting of the IRAS $60\,\mu m$ source counts from three major infrared emitters (starburst galxies,
   spiral galaxies and Seyferts). The source counts of starburst galaxies and Seyferts are from the Monte-Carlo simulation
   where the evolution of both activities are triggered by galaxy-galaxy interactions/mergers during structure
   formation. The spiral galaxy is assumed to have a mild constant star formation history, i.e. a non-evolving population in
   our calculation. The data are from the IRAS Point Source Catalogue (1985)(PSC), Hacking et al. IRAS deep survey (HCH), FSC from
   deep surveys by Moshir et al. (1992) and Saunders (1990). \label{fig4}}
   \end{figure}

   \begin{figure}
   \label{fig5}
   \includegraphics[height=8cm,width=8cm]{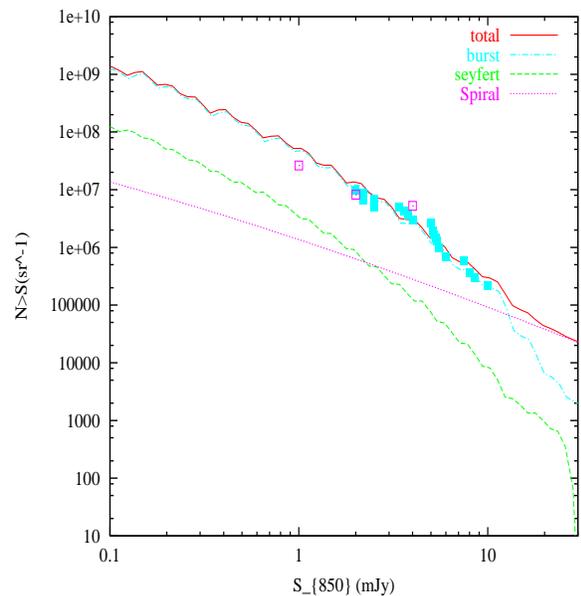}
   \caption{Integral number counts at $850\,\mu m$. The open squares are from Blain et al. (1999), and
   the filled squares are from Barger, Cowie and Sanders (1999). The meaning of lines are the same
   as in previous figures.
   \label{fig5}}
   \end{figure}

   \begin{figure}
   \label{fig6}
   \includegraphics[height=8cm,width=8cm]{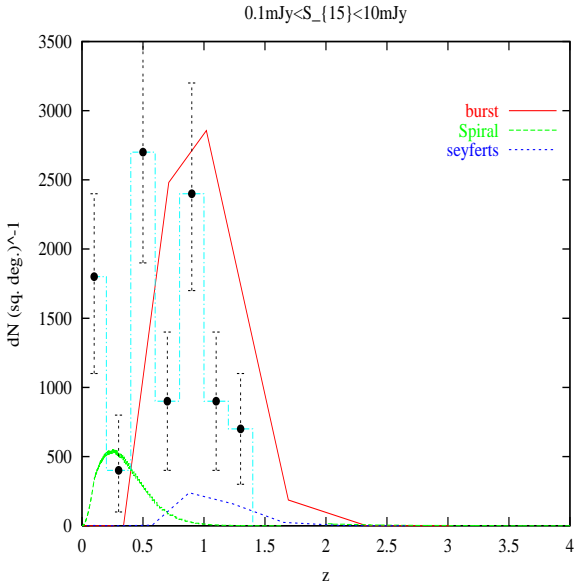}
   \caption{The redshift distribution of three infrared contributors (starburst galaxies, spiral galaxies and
   AGNs) at a flux range of $0.1\,mJy\sim\, 10\,mJy$ at $15\,\mu m$ from our model calculation.
   The redshift of these ISO far-infrared sources
   would cover a wide range and peak near $z\sim 1$. We include
   also the AGN contribution in our model based on the observed AGN Local
   Luminosity Function from Rush et al. (1993).
   \label{fig6}}
   \end{figure}

   \begin{figure}
   \label{fig7}
   \includegraphics[height=8cm,width=8cm]{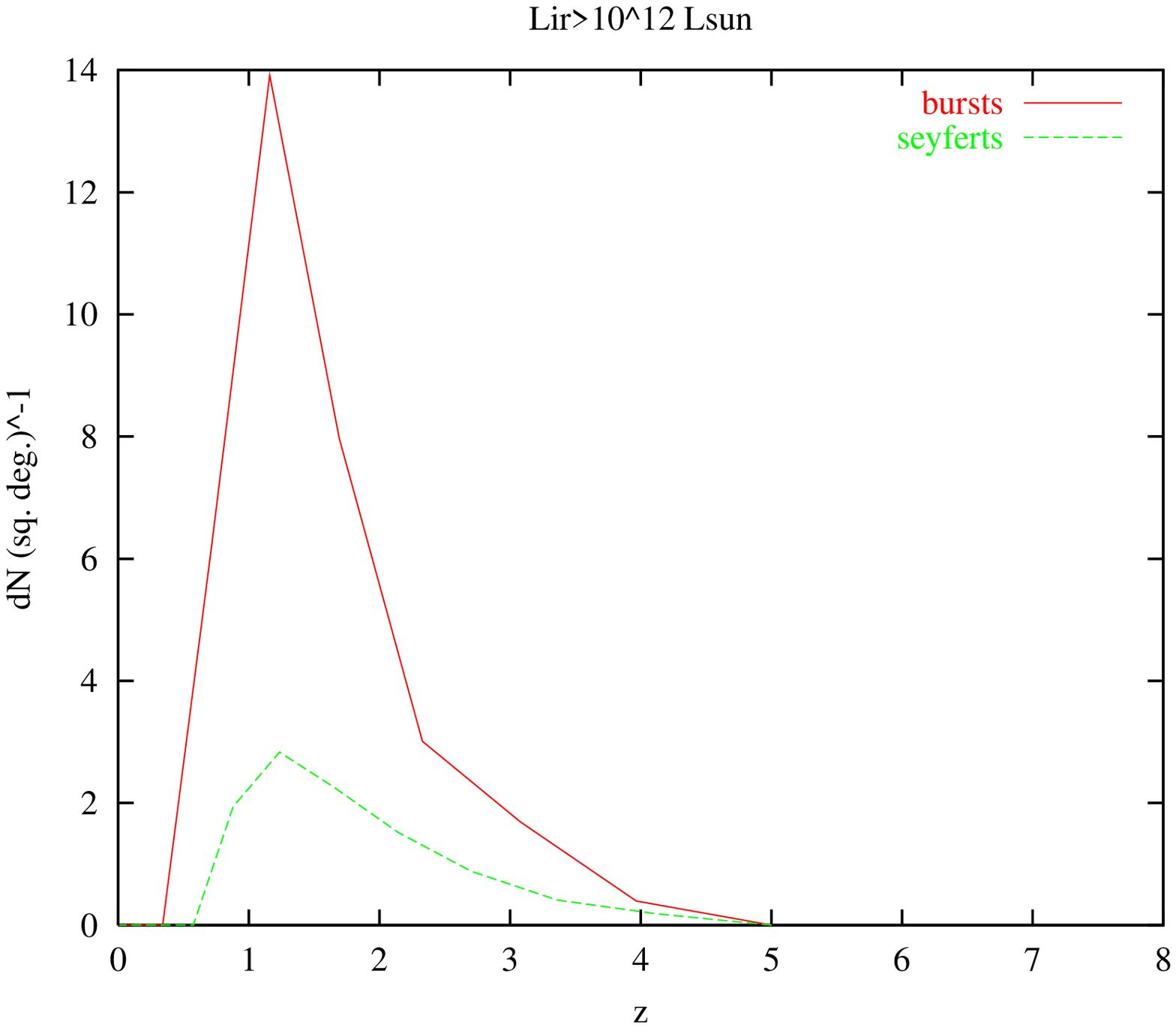}
   \caption{The redshift distribution of ultraluminous infrared sources (starbursts or AGNs)
   with $L_{ir}>10^{12}\,L_{\odot}$ from our model calculation.
   The number density of the ultraluminous infrared sources increase dramatically untill $z\sim 1$, but
   quickly decrease afterward.  \label{fig7}}
   \end{figure}

\section{Summary}
We have described a galaxy evolutionary scenario with galaxy
mergers in a CDM cosmology, where starburst and AGN activities may
be triggered by the merger events. With a reasonable assumption of
the ultraluminous infrared burst phase from gas rich mergers at
high redshift, we sucessfully interpreted the strong evolution of
IRAS 60$\,\mu m$ deep survey, leaving the infrared background
still with a low limit of $1.9\,nW\,m^{-2}\,sr^{-1}$, consistent
with the upper limits from recent TeV $\gamma$ ray detection of
nearby Blazars.

Lacking a comprehensive understanding of the gas and dust
environment of faint ISO sources, especially the starburst merging
system at $z\sim 1$, we adopt the template spectral energy
distribution such as that of Arp220 by Silva et al. (1998) as
typical for nearby starburst galaxies. Meanwhile, we construct a
simple SED for the starburst mergers at $z\sim 1$. The
far-infrared emission in such a system is modelled by a single
temperature, optically thin dust law with modified black body
emission. The MIR emission feature is assumed to be similar to Arp
220, but modified by the flux correlation from IRAS and ISOCAM
observations. In this case, we can further investigate such a
merger-driven galaxy evolutionary scenario at other infrared and
submillimeter wavelengths with ISO and SCUBA deep surveys. Our
calculation shows that the current results of multi-wavelength
deep surveys at ISOCAM 15$\mu m$, ELAIS 90$\mu m$, FIRBACK 170$\mu
m$, IRAS 60$\mu m$ and SCUBA 850$\mu m$ number counts could be
sufficiently accounted for by the merger-triggered infrared
enhancement at $z\sim 1$ from our model with the dust temperature
($T\sim 65\,K$, and $\beta \sim 1.5$), slightly higher than the
local starburst galaxy Arp220. Future accurate redshift
measurements and multiband photometries would provide a robust
model check.

The background levels at these wavelengths are estimated from our
model calculation, which gives 2.4$nW\,m^{-2}\,sr^{-1}$ at
15$\mu m$, 3.8$ nW\,m^{-2}\,sr^{-1}$ at
90$\mu m$, 10.6$nW\,m^{-2}\,sr^{-1}$ at 170$\mu m$, still compatible with the cosmic infrared
background level both from the upper limit of high energy TeV $\gamma$ ray detection of nearby Blazars
and from COBE and ISO results.

The redshift distribution of the luminous infrared sources within the
ISOCAM $15\,\mu m$ detection flux range ($0.1\,mJy\sim 10\,mJy$) from our calculation is plotted in Fig.~\ref{fig6}. The redshift
distribution of these sources cover a wide redshift range from $0.5\,\,\sim \,\, 2.5$ and
peak around a mean redshift of $z\sim 1$.

We also plot the redshift distribution of ULIGs ($\nu
L_{\nu}>10^{12}L_{\odot}$) in Fig.~\ref{fig7}. It shows a strong
increase in the ultraluminous infrared population untill a mean
redshift $z\sim 1$, and decreases by a factor of about 2 by $z\sim
2-3$. This is probably the major difference between our current
calculation and other models, and indicates that the infrared
luminous tail may be produced at the cosmic epoch of $z\sim 1$,
when the merger rate and the size of parent galaxies are suitable
for such an infrared emission enhancement.

A brief discussion about the fraction of contribution from AGNs
and starbursts from our calculation is given in section
\ref{discussion}. We assumed in the model a similar evolutionary
track for the starburst galaxies and AGNs based on the idea that
both AGNs and starbursts may be triggered by galaxy interactions,
where the AGN population is constrained by the observed Local
Luminosity Function of Seyferts from Rush et al. (1993). Given the
uncertainty of the dust properties of ULIGs, especially for those
harboring an AGN in the center, we adopt here only two typical SED
templates of Cloverleaf QSO and F10214+4724. In this case, we give
a rough estimation of the relative abundance of AGN and
starburst-powered ULIGs ($L_{ir}>10^{12}\,L_{\odot}$) of $\sim
1/5$, which seems to be close to the recent submillimeter
observations of Chandra X-ray sources (Almaini et al. 1999, Barger
et al. 2001, Gunn \& Shanks 2001).

\begin{acknowledgements}
    We acknowlege helpful discussions with Profs. P. L. Biermann, M. Harwit, N. Arimoto, and Drs. K. Kawara, Y. Taniguchi, T. Yamada.
    YPW would like to thank the anonymous referee for the kind suggestions and comments, which helped to improve the paper significantly.
    YPW is supported by NSFC 10173025 and Chinese post-doctor science foundation. YPW is also
    very grateful for the hospitality of NAOJ staff and COE fellowship of Japan, where the final stage of this work was finished.
\end{acknowledgements}

\end{document}